\documentclass[superscriptaddress,aps,pra,amsfonts,notitlepage,onecolumn,nofootinbib]{revtex4-2}

\usepackage{xcolor,graphicx}
\usepackage{breqn,amsmath,amssymb}
\usepackage{mathtools}
\usepackage{enumerate}

\usepackage{booktabs}
\usepackage{makecell}

\usepackage{comment}

\usepackage{subcaption}

\usepackage[T1]{fontenc}
\usepackage{natbib,hyperref}

\usepackage{mathtools}

\usepackage{centernot}

\usepackage[utf8]{inputenc}

\definecolor{linkcolor}{HTML}{799B03}
\definecolor{urlcolor}{HTML}{799B03}
\usepackage[english]{babel}
\usepackage{amsmath, amsthm, amsfonts, amssymb,amscd}
\usepackage[all]{xy}
\usepackage{graphicx}
\usepackage{import}
\usepackage{xifthen}
\usepackage{float}
\usepackage{transparent}
\usepackage{multirow}
\usepackage{cellspace}
\usepackage{colortbl}
\usepackage{bbm}
\usepackage{mathrsfs}
\usepackage{csquotes}
\usepackage{natbib}

\graphicspath{{pic/}}
\def\[{\begin{equation}}
\def\]{\end{equation}}

\DeclareUnicodeCharacter{0301}{`}

\makeatletter
\let\cat@comma@active\@empty
\makeatother

\usepackage{IEEEtrantools}

\begin{document}

%
\title{Ghost-free, gauge invariant SVT generalizations of Horndeski theory}

\author{S. Mironov}
\email{sa.mironov\_1@physics.msu.ru}
\affiliation{Institute for Nuclear Research of the Russian Academy of Sciences, 117312 Moscow}
\affiliation{Institute for Theoretical and Mathematical Physics,
MSU, 119991 Moscow, Russia}
\affiliation{NRC, "Kurchatov Institute", 123182, Moscow, Russia}

\author{A. Shtennikova}
\email{shtennikova@inr.ru}
\affiliation{Institute for Nuclear Research of the Russian Academy of Sciences, 117312 Moscow}
\affiliation{Institute for Theoretical and Mathematical Physics,
MSU, 119991 Moscow, Russia}

\author{M. Valencia-Villegas}
\email{mvalenciavillegas@itmp.msu.ru}
\affiliation{Institute for Theoretical and Mathematical Physics,
MSU, 119991 Moscow, Russia}
\affiliation{Institute for Nuclear Research of the Russian Academy of Sciences, 117312 Moscow}

\begin{abstract}
We analyze the generalizations of Kaluza--Klein compactifications of 5D Horndeski theory. They are Scalar--Vector--Tensor (SVT) theories with higher derivatives in the action, but with second order equations of motion. The vector field is invariant under a U(1) gauge transformation and the Scalar--Tensor sector corresponds to Horndeski theory. A subclass of these SVT theories is such that the Horndeski functions  $G_4(\pi,X)$ and $G_5(\pi)$ remain free, while the speed of the tensor and vector modes is exactly the same. We show a subclass where the vector sector retains freedom through new functions of $\pi,\, X$ while the speed of the vector modes still tracks the speed of the tensor modes. 
\end{abstract}

\maketitle
\tableofcontents

\section{Introduction}

A large portion of modified gravity theories for early and late time cosmology are built with scalar fields in the action, besides the metric. If we consider up to second derivatives of such scalar field in the Lagrangian, but keeping second order equations of motion ---besides assumptions on vanishing Torsion and nonmetricity--- one is led to the unique and quite generic Horndeski theory, later rediscovered as generalized Galileons \cite{horndeski1974second,nicolis2009galileon,Deffayet:2011gz,Deffayet:2009wt}. 

This theory contains nonminimal couplings of the scalar to curvature, meaning that generally the speed of the tensor modes is different from unity. This property  has been thought to be a problem, because the speed of gravity is nearly the same to the speed of light \cite{LIGOScientific:2017vwq,Abdalla:2022yfr}. Thus, this led to conclude that pure Scalar--Tensor theories are highly constrained to have few freedom in the nonminimal couplings. For instance, the non-minimal couplings with derivatives in Horndeski theory were thought to be ruled out \cite{Bettoni:2016mij,Ezquiaga:2017ekz,Sakstein:2017xjx,Baker:2017hug,Creminelli:2017sry}. Which is unfortunate given that nonminimal couplings ---as a potential form of dynamical dark energy--- are receiving new attention in relation to the Baryonic Acoustic Oscillations data from DESI \cite{Ye:2024ywg,Wolf:2024eph,DESI:2024mwx,DESI:2025fii,Giare:2024gpk}. 

However, in the line of thought proposed in \cite{Mironov:2024idn}, it is possible to evade these constraints on Horndeski theory if one also considers a U(1) gauge vector field in the action, and a specific coupling of the vector to the scalar modification of gravity. Namely, a coupling with the property that the vector modes propagate at the same speed as gravitational waves in Horndeski theory. This specific Scalar--Vector--Tensor theory (S--V--T), with the Horndeski action in the S--T sector, has a clear motivation if one identifies the vector modes with the photon propagating in the cosmological medium. Indeed, it aligns with the experimental constraint on the (near) equality of the speed of gravitational waves ($c_g$)  and light ($c_A$), $\vert {c_g}/{c_A}-1 \vert \leq 5 \times 10^{-16}$\cite{LIGOScientific:2017vwq,Abdalla:2022yfr} . Following  \cite{Mironov:2024idn,Mironov:2024umy,Mironov:2024wbx} we refer to them in this work as ``Luminal SVT generalizations of Horndeski theories'', because the graviton is automatically ``luminal'' {\it i.e.} it propagates at exactly the same speed of the modified photon --- without choosing particular scalar potentials in the Horndeski action.  Besides, similar scalar--vector couplings for theories beyond Horndeski were shown to widen the phenomenologically viable classes, even beyond these speed test constraints \cite{Mironov:2024wbx}.

Cosmological studies with fundamental vector fields have been more widely explored in inflation, dark energy models or, for instance, in the context of primordial magnetic fields, just to name a few applications \cite{Horndeski:1976gi,Esposito-Farese:2009wbc,Golovnev:2008cf,BeltranJimenez:2008iye,BeltranJimenez:2013btb,Heisenberg:2018mxx,Tasinato:2013oja}. Indeed, recent efforts to build very general Einstein--Maxwell theories \cite{Colleaux:2023cqu,Colleaux:2024ndy,Colleaux:2025vtm} reveal the relevance of mapping U(1) gauge--invariant SVT theories for potential future applications, such as in \cite{Heisenberg:2018acv} \footnote{Simultaneously with this manuscript, another work \cite{Gorji:2025lqr} appeared on linear Einstein--Maxwell--Scalar theories.}. The task is in fact not simple, as we are told by the no-go theorem for U(1) gauge invariant vector Galileons on flat spacetime (without scalars)--- with higher derivatives of a vector and the metric in the action, but with second order equations of motion \cite{Deffayet:2013tca}. This essential difficulty is the reason why a large part of SVT theories in the literature have been built breaking the U(1) gauge invariance \cite{Heisenberg:2014rta,Heisenberg:2018acv,Petrov:2018xtx}, and also a reason why the new method to construct U(1) gauge invariant  SVT theories in four dimensions ---by Kaluza--Klein compactifications of higher dimensional S--T theories \cite{Mironov:2024idn,Mironov:2024wbx}--- is relevant. 

 In this work we first generalize the SVT theory (in 4 dimensions) which is obtained by a dimensional reduction of 5 dimensional Horndeski theory, shown in \cite{Mironov:2024umy}. In other words, we show that the latter theory is just a particular case of a broad U(1) gauge invariant SVT theory with new free functions, such that the equations remain of second order. 
 
 Next, we present a family of "Luminal SVT generalizations of Horndeski theories" with additional free functions and with similar properties as the Kaluza--Klein reduction of Horndeski theory shown in \cite{Mironov:2024idn}. Namely, the speed of the tensor and vector modes is exactly  the same on any cosmological background, even for the general $G_4(\pi,X)$ and $G_5(\pi)$ functions in the Horndeski action. Nevertheless, the luminal SVT theories shown in this work retain freedom for independent modifications of the vector and tensor sectors.

This paper is structured as follows. First we construct the SVT models in section \ref{sec const SVT}. We give the assumptions and the basis of SVT couplings in \ref{sec basis}. Then, we show the SVT theories that have second order equations of motion in all of the fields \ref{sec ghost-freeSVT}. We summarize the model in section \ref{sec summaryModelWithHorndeski}. Then we consider cosmological perturbations of these SVT theories in \ref{sec perturbations}. We introduce notation in \ref{sec notation perts} and then give the actions at quadratic order in \ref{sec quadratic actions}. We end in \ref{sec luminalSVT ext} singling out the subclass of SVT theories that have vector and tensor modes of Horndeski that propagate at exactly the same speed. We conclude in section \ref{sec conclusions}.

\section{SVT generalizations of Horndeski theory}\label{sec const SVT}

 The first objective in this paper is to extend the theories beyond the particular SVT that can be obtained by a Kaluza--Klein reduction of 5D Horndeski theory, which was shown in part in \cite{Mironov:2024idn} and completely in \cite{Mironov:2024umy}. Let us refer to it as KK-SVT. In other words, we wish to construct a family of SVT theories that contains the latter just as a special case. 
 
 Keeping a family connected to the KK-SVT is physically compelling, because it contains Horndeski in the Scalar--Tensor sector, furthermore, the vector is high in derivatives while also being invariant under U(1) gauge transformation.  Finally, it has no Ostrogradski ghosts, and  it has a subclass where the tensor and vector modes propagate at exactly the same speed. What we will show below is that a larger family of SVT theories shares all of these physically interesting properties---at least on any cosmological background--- even if they cannot be obtained by a Kaluza-Klein reduction of 5D Horndeski theory.

 \subsection{The pure SVT part: basis of couplings to the $U(1)$ vector}\label{sec basis}

 In this section we take a basis of independent terms for the Lagrangian of the scalar-vector-tensor theory. First of all, we start with the action that was obtained by a Kaluza-Klein dimensional reduction of Horndeski theory \cite{Mironov:2024idn,Mironov:2024umy}. Following \cite{Mironov:2024umy}, after a number of integrations by parts and simplifications, this Lagrangian can be expressed in the following form: we take a basis of Scalar--Vector--Tensor couplings $V_i$, with $i=1,\dots 33$, each of which will be multiplied by a coefficient $\lambda_i(\pi,X)$ in the Lagrangian, where $\pi$ is the real scalar field and $X=\partial_\mu \pi \partial^\mu \pi$ its first derivatives. They are given in equations (\ref{eqn SVTbasis}) . We take a ($+,-,-,-$) signature for the metric. The pure SVT Lagrangian ---namely, the part that necessarily has the vector field coupled to the scalar and the metric--- takes the form
\begin{align}
    \mathcal{L}_{SVT_{basis}}=\sum^{33}_{i=1} \lambda_i(\pi,X)\,\,V_i\,,\label{eqn lagSVTa}
\end{align} 
with,
    \begin{align}
    V_{1}&=F_{\alpha \beta } F^{\alpha \beta }\,,& V_{2}&=F_{\alpha \beta } F^{\alpha \beta } (\nabla_{\gamma }\nabla^{\gamma }\pi\,)\,,& \nonumber\\[1.5ex]
    V_{3}&=F^{\alpha }{}_{\epsilon } \nabla_{\nu}\pi\, \nabla^{\nu}\nabla_{\mu}\pi\, \nabla_{\alpha }F^{\epsilon \mu}\,, & V_{4}& =F_{\delta \epsilon } \nabla_{\nu}\pi\, \nabla^{\nu}\nabla^{\alpha }\pi\, \nabla_{\alpha }F^{\delta \epsilon }\,,& \nonumber \\[1.5ex]
    V_{5}&=F_{\alpha \beta } F^{\alpha }{}_{\delta } \nabla^{\beta }\pi\, \nabla^{\delta }\pi\, \,,& V_{6}&=F_{\alpha \beta } F^{\alpha }{}_{\delta } \nabla^{\delta }\nabla^{\beta }\pi\, \,, & \nonumber \\[1.5ex]
    V_{7}&=F_{\alpha \beta } F^{\alpha \beta } \nabla_{\nu}\nabla^{\mu}\pi\, \nabla^{\nu}\pi\, \nabla_{\mu}\pi\, \,,& V_{8}&=F_{\alpha \beta } F^{\alpha \beta } F_{\epsilon \nu} F^{\epsilon }{}_{\lambda } \nabla^{\nu}\pi\, \nabla^{\lambda }\pi\, \,,\nonumber \\[1.5ex]
    V_{9}&=F_{\alpha \beta } F^{\alpha }{}_{\delta } F^{\beta }{}_{\nu} F^{\delta }{}_{\lambda } \nabla^{\nu}\pi\, \nabla^{\lambda }\pi\, \,, & V_{10}&=F_{\alpha \beta } F^{\alpha \beta } R\,, & \nonumber \\[1.5ex]
    V_{11}&=(F_{\alpha \beta } F^{\alpha \beta })^2\,,& V_{12}&=F_{\alpha \beta } F^{\alpha }{}_{\delta } F^{\beta }{}_{\nu} F^{\delta \nu}\,, & \nonumber \\[1.5ex]
    V_{13} &= F_{\alpha \beta } F^{\alpha \beta } (\nabla_{\gamma }\nabla^{\gamma }\pi\,)^2\,, 
	&
	V_{14} &= F_{\alpha \beta } F^{\alpha}{}_{\delta } R \nabla^{\beta }\pi\, \nabla^{\delta }\pi\, \,, \nonumber\\[1.5ex]
	V_{15} &= F_{\alpha \beta } F^{\alpha \beta } \nabla_{\eta }\nabla_{\mu}\pi\, \nabla^{\eta }\nabla^{\mu}\pi\,,
	&
	V_{16} &= F_{\alpha \beta } F^{\alpha}{}_{\delta } \nabla_{\nu}\nabla^{\beta }\pi\, \nabla^{\nu}\pi\, \nabla^{\delta }\pi\,, \label{eqn SVTbasis}\nonumber \\[1.5ex]
	V_{17} &= F_{\alpha \beta } F^{\alpha}{}_{\delta } (\nabla_{\gamma }\nabla^{\gamma }\pi\,) \nabla^{\delta }\nabla^{\beta }\pi\,,
	&
	V_{18} &= F_{\delta \epsilon } (\nabla_{\gamma }\nabla^{\gamma }\pi\,) \nabla_{\alpha }F^{\delta \alpha } \nabla^{\epsilon }\pi\, \,, \nonumber\\[1.5ex]
	V_{19} &= F_{\alpha \beta } F^{\alpha}{}_{\delta } \nabla_{\nu}\nabla^{\beta }\pi\, \nabla^{\nu}\nabla^{\delta }\pi\,,
	&
	V_{20} &= F_{\alpha \beta } F_{\gamma \delta } R^{\alpha \gamma } \nabla^{\beta }\pi\, \nabla^{\delta }\pi\, \,, \nonumber\\[1.5ex]
	V_{21} &= F_{\delta \epsilon } \nabla_{\alpha }F^{\delta \alpha } \nabla^{\epsilon }\pi\, \,,
	&
	V_{22} &= - F_{\delta \epsilon } \nabla_{\alpha }F^{\alpha \mu} \nabla^{\delta }\pi\, \nabla^{\epsilon }\nabla_{\mu}\pi\, \,,
 \\[1.5ex]
	V_{23} &= F_{\alpha \beta } F_{\gamma \delta } \nabla^{\gamma }\nabla^{\alpha }\pi\, \nabla^{\delta }\nabla^{\beta }\pi\,,
	&
	V_{24} &= F_{\delta \epsilon } \nabla_{\alpha }F^{\delta \eta } \nabla^{\epsilon }\pi\, \nabla_{\eta }\nabla^{\alpha }\pi\, \,, \nonumber\\[1.5ex]
	V_{25} &= F_{\delta \epsilon } \nabla_{\nu}\pi\, \nabla^{\nu}\nabla^{\epsilon }\pi\, \nabla_{\alpha }F^{\delta \alpha }\,,
	&
	V_{26} &= F_{\alpha \beta } F^{\alpha}{}_{\delta } (\nabla_{\gamma }\nabla^{\gamma }\pi\,) \nabla^{\beta }\pi\, \nabla^{\delta }\pi\,, \nonumber\\[1.5ex]
	V_{27} &= F_{\alpha \beta } F_{\gamma \delta } \nabla^{\beta }\pi\, \nabla^{\gamma }\nabla^{\alpha }\pi\, \nabla^{\delta }\pi\, \,,
	&
	V_{28} &= F_{\alpha \beta } F^{\alpha}{}_{\delta } R^{\beta \delta }\,, \nonumber\\[1.5ex]
	V_{29} &= F_{\alpha \beta } F_{\gamma \delta } R^{\alpha \gamma \beta \delta }\,,
	&
	V_{30} &= F_{\alpha \beta } F^{\alpha}{}_{\delta } (\nabla_{\gamma }\nabla^{\gamma }\pi\,)^2 \nabla^{\beta }\pi\, \nabla^{\delta }\pi\,, \nonumber\\[1.5ex]
	V_{31} &= F_{\alpha \beta } F^{\alpha}{}_{\delta } \nabla^{\beta }\pi\, \nabla^{\delta }\pi\, \nabla_{\eta }\nabla_{\mu}\pi\, \nabla^{\eta }\nabla^{\mu}\pi\,,
	&
	V_{32} &= - F_{\alpha \beta } F_{\gamma \delta } \nabla_{\nu}\nabla^{\alpha }\pi\, \nabla^{\nu}\nabla^{\gamma }\pi\, \nabla^{\beta }\pi\, \nabla^{\delta }\pi\, \,, \nonumber\\[1.5ex]
	V_{33} &= F_{\alpha \beta } F_{\gamma \delta } (\nabla_{\gamma }\nabla^{\gamma }\pi\,) \nabla^{\beta }\pi\, \nabla^{\gamma }\nabla^{\alpha }\pi\, \nabla^{\delta }\pi\,.\nonumber
\end{align}
\vspace{0.3cm}

Where $F_{\mu\nu}=\nabla_\mu A_\nu-\nabla_\nu A_\mu$ is the usual field strength, where $\nabla$ is the torsionless, metric compatible covariant derivative. Note that  we consider the  basis $V_i$ with couplings  of $F$ to up to the square of second derivatives of the scalar $\pi$, and to first power of the Riemann tensor. Because $V_i$ are multiplied by generally independent functions of $\pi$ and $X$ in the Lagrangian, $\lambda_i(\pi,X)$, each block $\lambda_{j^*}(\pi,X)V_{j^*}$, for fixed $j^*$ (not summed over $j$) is in principle fully independent of the other blocks and cannot be reduced by integration by parts. It is clear that we do not aim for generality. Instead we wish to expand the family of KK-SVT, and remain connected to Horndeski theory and its luminal SVT subclass.

This basis  is  such that it reduces under a very precise choice of functions $\lambda_i(\pi,X)=\lambda_i(G_4,\,G_5,G_6)$, which we show below, to the KK--SVT lagrangian derived in \cite{Mironov:2024umy} for the metric, the vector and a single scalar field (namely, with Galileon field but without dilaton of  \cite{Mironov:2024umy}).

\subsection{The ghost-free U(1) gauge invariant SVT}\label{sec ghost-freeSVT}

It is also clear that the Lagrangian (\ref{eqn lagSVTa}) for general $\lambda_i$ would yield fourth order equations of motion. In this section we first find the largest set of free functions $\lambda_k$ that yields second order equations of motion. For concreteness, keeping the same numbering, let us choose as free general functions the first $\lambda_k$'s. Any other choice of initial functions, would of course lead to the same physics, just with re-labeled free functions. Requiring the equations of motion to be of second order, we obtain the expression of 21 $\lambda_i$ in terms of the 12 free functions and their derivatives. The following are self-consistent, free of ghosts and $U(1)$ gauge invariant Lagrangians: with the notation  $\lambda_{k,\,X}=\frac{\partial \lambda_k}{\partial X}$,
\begin{subequations}
    \begin{align}
    L_{1}&=V_{1}  \lambda_{1}\,,\label{eqn lagSVTFinal0}\\ & \nonumber \\
L_{2}&=(V_{2}+4 V_{21}) \lambda_{2}+(4 V_{26}-4 V_{27})\lambda_{2X}\,,\\ & \nonumber \\
L_{3}&=\Bigg(\frac{V_{14}}{2}-V_{17}-2 V_{18}+V_{19}-V_{20}-2 V_{22}+V_{23}+2 V_{24}+V_{25}+V_{3}\Bigg) \lambda_{3}\nonumber\\&
+(-V_{30}+V_{31}+2 V_{32}+2 V_{33}) \lambda_{3X}+(-V_{26}+V_{27}+V_{16}) \lambda_{3\pi}\,,\\ & \nonumber \\
L_{4}&=(-V_{14}+2 V_{17}+4 V_{18}-2 V_{19}+2 V_{20}+4 V_{22}-2 V_{23}-4 V_{24}-2 V_{25}+V_{4}) \lambda_{4}\nonumber\\&
+(2 V_{30}-2 V_{31}-4 V_{32}-4 V_{33}) \lambda_{4X}+(2 V_{26}-2 V_{27}-2 V_{16}) \lambda_{4\pi}\,,\\ & \nonumber \\
L_{6}&=(V_{6}+2 V_{21}) \lambda_{6}+(2 V_{26}-2 V_{27}) \lambda_{6X}\,,\\ & \nonumber \\
L_{7}&=(2 V_{26}-2 V_{27}-4 V_{16}+V_{7}) \lambda_{7}\,,\\ & \nonumber \\
L_{10}&=(-4 V_{28}+2 V_{29}+V_{10}) \lambda_{10}+(-2 V_{13}+2 V_{15}+8 V_{17}-8 V_{19}-4 V_{23}) \lambda_{10X}\,\label{eqn lagSVTFinalF}
\end{align}
\end{subequations}
and also $L_{k^*}$ for ${k^*} \in \{5,\,8,\,9,\,11,\,12\}$, which are low in derivatives, but they are needed if we want that  this basis of Lagrangians contains the KK-SVT theory as a special case \cite{Mironov:2024idn}. The latter are simply written as $L_{k^*}=V_{k^*} \,\lambda_{k^*}(\pi,X)$ for fixed ${k^*}$ (not summed index). In particular, note that $L_1$ is also low in derivatives and it contains the Maxwell Lagrangian as a special case when $\lambda_1$ is a constant. Furthermore, we can note that in this basis $L_4$ is redundant. Thus we can set $\lambda_4=0$ and redefine $\lambda_3$ to reproduce the KK reduction of 5D Horndeski theory \cite{Mironov:2024idn}. For instance, we do not consider $\lambda_4$ in the next section. Let us also note that, by construction, this basis of ghost-free Lagrangians  contain --- only as a special  case --- the SVT where the tensor and vector modes propagate at exactly the same speed. This is further discussed in section \ref{sec relation to KKSVT}.

For simplicity, let us also explicitly give the relations between $\lambda_i$ in the basis (\ref{eqn SVTbasis}) stated above: namely, the basis of ghost-free SVT lagrangians can be obtained from (\ref{eqn lagSVTa}) with the following relations

\begin{equation}
\begin{array}{cc}
\lambda_{26} = 2\,\lambda_{6X}+4\,\lambda_{2X}+2\,\lambda_{4\pi}-\,\lambda_{3\pi}+2\,\lambda_{7} \,, \quad &
\lambda_{27} = -2\,\lambda_{6X}- 4\,\lambda_{2X} -2\,\lambda_{4\pi} +\lambda_{3\pi}- 2\,\lambda_{7} \, , \\ & \\
\lambda_{16} = -2\,\lambda_{4\pi}+\lambda_{3\pi}-4\,\lambda_{7} \,,&
\lambda_{21} = 2\lambda_{6} + 4\lambda_{2} \,, \\ & \\
\lambda_{30} = -\frac{1}{2}\lambda_{24X} \,,&
\lambda_{31} = \frac{1}{2}\lambda_{24X} \,, \\ & \\
\lambda_{32} = \lambda_{24X} \,,&
\lambda_{13} = -2\lambda_{10X} \,, \\ & \\
\lambda_{14} = \frac{1}{4}\lambda_{24} \,,&
\lambda_{15} = 2\lambda_{10X} \,, \\ & \\
\lambda_{17} = 2\,\lambda_{4}-\,\lambda_{3}+8\,\lambda_{10X} \,,&
\lambda_{18} = -\lambda_{24} \,, \\ & \\
\lambda_{19} = -8\,\lambda_{10X}-2\,\lambda_{4}+\lambda_{3} \,,&
\lambda_{20} = -\frac{1}{2}\lambda_{24} \,, \\ & \\
\lambda_{22} = 4\lambda_{4} - 2\lambda_{3} \,,&
\lambda_{23} = -2\,\lambda_{4}+\lambda_{3}-4\,\lambda_{10X}  \\ & \\
\lambda_{24} = -4\,\lambda_{4}+2\,\lambda_{3} \,,&
\lambda_{25} = -2\,\lambda_{4}+\lambda_{3} \,, \\ & \\
\lambda_{28} = -4\lambda_{10} \,,&
\lambda_{29} = 2\lambda_{10} \,, \\ & \\
\lambda_{33} = \lambda_{24X} \,.&
\end{array}\label{eqn noGhostRel}
\end{equation}

\subsection{The model: SVT generalization of Horndeski theory}\label{sec summaryModelWithHorndeski}

Let us note that the Scalar--Vector--Tensor U(1) gauge invariant theories ${L}_{k}$ shown above cannot replace S--T theories like Horndeski for most cosmological applications. Indeed, a quick inspection shows that all SVT couplings in (\ref{eqn SVTbasis}) are at least quadratic in $F$ or derivatives of $F$. Thus, at first order in a perturbative expansion the scalar and tensor perturbations will be multiplied by the background vector field, which in a FLRW cosmology must vanish. In other words, at quadratic order in the action, the ${L}_{k}$ theories only have an in principle non--trivial vector sector of their own. The contribution of scalar and tensor perturbations will be non trivial only at second order in a perturbative expansion or on more general, non--isotropic backgrounds, such as in regions of strong magnetic fields.

Clearly, for modifications of gravity we need the tensor modes at first order in the expansion, thus we add the most general Scalar--Tensor sectors (with one scalar) built under the same principle used to construct $\mathcal{L}_{SVT}$, namely, keeping second order equations of motion. Thus, by the standard theorem of Horndeski, it is clear that the action takes the form,
\begin{align}
S=\int&\mathrm{d}^4x\sqrt{-g}\,\mathcal{L}_{HSVT},\\
\nonumber\\
\mathcal{L}_{HSVT}&=\mathcal{L}_H+\mathcal{L}_{SVT}\,,\label{eqn fullL}
\end{align}
where $\mathcal{L}_{SVT}$ is given by 
\begin{align}
    \mathcal{L}_{SVT}=\sum_{k} L_k  \,,\label{eqn lagSVTFinal}
\end{align}
with $k \in {1-3,\,5-12}$ and $\mathcal{L}_H$ is the Horndeski action (generalized Galileons) in 4 dimensions, written with four general functions  $F,\, K,\, G_4,\, G_5$ of the scalar field $\pi$ and $X$, as usual \cite{horndeski1974second,Deffayet:2011gz,Kobayashi:2011nu},
\begin{subequations}
    \label{lagrangian}
    \[\mathcal{L}_H=\mathcal{L}_2 + \mathcal{L}_3 + \mathcal{L}_4 + \mathcal{L}_5\]\label{eqn Horndeski}
    \vspace{-0.8cm}
    \begin{align}
    &\mathcal{L}_2=G_2(\pi,X),\\
    &\mathcal{L}_3=G_3(\pi,X)\Box\pi,\\
    &\mathcal{L}_4=-G_4(\pi,X)\,R+2G_{4X}(\pi,X)\left[\left(\Box\pi\right)^2-\pi_{;\mu\nu}\pi^{;\mu\nu}\right],\\
    &\mathcal{L}_5=G_5(\pi,X)G^{\mu\nu}\pi_{;\mu\nu}+\frac{1}{3}G_{5X}(\pi,X)\left[\left(\Box\pi\right)^3-3\Box\pi\,\pi_{;\mu\nu}\pi^{;\mu\nu}+2\pi_{;\mu \nu}\pi^{;\mu \rho}\pi_{;\rho}^{\;\;\nu}\right],&
    \end{align}
\end{subequations}
In the section below we analyze the cosmological perturbations  of the $\mathcal{L}_{HSVT}$ theory (\ref{eqn fullL}).\\

It is worth mentioning that one could also consider beyond Horndeski or even DHOST Lagrangian instead of $\mathcal{L}_{H}$, that would lead to higher order equations in scalar--tensor sector, but still keep the theory healthy without Ostrogradski ghost.

\subsection{A special case: The Kaluza--Klein reduction of 5D Horndeski}\label{sec relation to KKSVT}

As expected, by construction, the equations (\ref{eqn noGhostRel}) have as a special case of a theory without ghosts the Kaluza--Klein reduction of 5D Horndeski theory (without dilaton) \cite{Mironov:2024idn}. The latter can be obtained by taking,

\begin{equation}
\begin{array}{ccc}
\lambda_{1} =  - \dfrac{1}{4}\,{G_{4}} - \dfrac{15}{32}\,G_{6\pi \pi} X, & \lambda_{2} =  - \dfrac{1}{8}\,{G_{5}} - \dfrac{15}{32}\,G_{6\pi}, & \lambda_{3} = \dfrac{3}{4}\,G_{6X}, \\[2ex]
\lambda_{4} = \dfrac{15}{8}\,G_{6X}, & \lambda_{5} = G_{4X}+\dfrac{3}{8}\,G_{6\pi \pi}, & \lambda_{6} = \dfrac{1}{2}\,{G_{5}}+\dfrac{3}{8}\,G_{6\pi}, \\[2ex]
\lambda_{7} =  - \dfrac{15}{16}\,G_{6\pi X}, & \lambda_{8} =  - \dfrac{9}{8}\,G_{6X}, & \lambda_{9} = \dfrac{9}{4}\,G_{6X}, \\[2ex]
\lambda_{10} = \dfrac{3}{8}\,{G_{6}}, & \lambda_{11} = \dfrac{9}{64}\,{G_{6}}, & \lambda_{12} =  - \dfrac{9}{32}\,{G_{6}}, \\[2ex]
\lambda_{13} =  - \dfrac{3}{4}\,G_{6X}, & \lambda_{14} =  - \dfrac{3}{2}\,G_{6X}, & \lambda_{15} = \dfrac{3}{4}\,G_{6X}, \\[2ex]
\lambda_{16} = \dfrac{3}{4}\,G_{6\pi X}, & \lambda_{17} = 6\,G_{6X}, & \lambda_{18} = 6\,G_{6X},\\[2ex]
\lambda_{19} = -6\,G_{6X}, & \lambda_{20} = 3\,G_{6X}, & \lambda_{21} = \dfrac{1}{2}\,{G_{5}} - \dfrac{9}{8}\,G_{6\pi}, \\[2ex]
\lambda_{22} = 6\,G_{6X}, & \lambda_{23} =  - \dfrac{9}{2}\,G_{6X}, & \lambda_{24} = -6\,G_{6X}, \\[2ex]
\lambda_{25} = -3\,G_{6X}, & \lambda_{26} = \dfrac{1}{2}\,G_{5X}, & \lambda_{27} =  - \dfrac{1}{2}\,G_{5X},\\[2ex]
\lambda_{28} =  - \dfrac{3}{2}\,{G_{6}}, & \lambda_{29} = \dfrac{3}{4}\,{G_{6}}, & \lambda_{30} = 3\,G_{6XX}, \\[2ex]
\lambda_{31} = -3\,G_{6XX}, & \lambda_{32} = -6\,G_{6XX}, & \lambda_{33} = -6\,G_{6XX}\,,
\end{array}
\end{equation}

such that $G_6$ is the only free function of that vector-scalar Galileon, besides the usual $G_2,\,G_3,\,G_4$ and $G_5$ of the Scalar--Tensor part.


\section{Cosmological perturbations of SVT theories}\label{sec perturbations}

In this Section we derive the quadratic action for perturbations above a spatially flat FLRW background.

\subsection{Notation: decomposition into irreducible components}\label{sec notation perts}
As usual, let us consider the perturbed metric and 4-vector as
\begin{eqnarray}
    ds^2&=&(g_{\mu\nu}+ h_{\mu\nu})\text{d}x^{\mu}\text{d}x^{\nu}\,,\label{eqn perturbed metric}\\
    A^{\mu}&=&A^{(0)\,\mu}+ \delta A^{\mu}\,, \label{eqn perturbed 4vector}
\end{eqnarray}
while the Galileon $\pi(x^{\mu})$ is expanded as $\pi(t)+\delta \pi(x^{\mu})$ and $\pi$ will be understood as background field,  or not, depending on the context. Isotropy automatically requires the gauge vector to have trivial background $A^{(0)\,\mu}=0$. The background, spatially flat FLRW metric is written as usual
\[ds^2=g_{\mu\nu}\text{d}x^{\mu}\text{d}x^{\nu} = dt^2 - a(t)^2 \left(dx^2 + dy^2 + dz^2\right)\,, \]
where $a(t)$ is the scale factor. 

We take the usual decomposition of perturbations $ h_{\mu \nu},\, \delta A^{\mu}$ into irreducible components under spatial rotations, as
\begin{subequations}
\begin{align}
&\delta h_{00}=2\Phi \\
&\delta h_{0 i}=-\partial_{i} \beta + Z_i, \\
&\delta h_{i j}=-2\Psi \delta_{i j}-2 \partial_{i}\partial_{j} E - \left(\partial_{i}W_j+\partial_{j}W_i\right)+h_{i j},\\
&\delta A_{0}=\gamma,\\
&\delta A_{i}=\partial_{i}\alpha + A_{i},\\
&\delta\pi=\chi, 
\end{align}    \label{eqn decomposition}
\end{subequations}
where $\Phi, \beta, \Psi, E, \chi, \varphi, \alpha, \gamma$ are scalar fields, $Z_{i}, W_{i}, A_{i}$ are transverse two-component vector fields and $h_{ij}$ is the transverse traceless two-component tensor. Let us stress that the we denote the transverse perturbation of $\delta A_i$ simply by $A_i$ in linearized expressions, with no risk of confusion with the full field $A^\mu$, because their use depends on the context. Similarly for $h_{\mu\nu}$ and the traceless transverse part $h_{ij}$.

Now, by the same argument as in the last section, because all SVT couplings in (\ref{eqn SVTbasis}) are at least quadratic in $F$, all the contributions to the background equations from the $\mathcal{L}_{SVT}$ will be multiplied by $A^{(0)\,\mu}=0$. Thus, the background equations take the exactly same form as in Horndeski theory (See for instance \cite{Kobayashi:2019hrl}).

\subsection{Quadratic actions on the FLRW background}\label{sec quadratic actions}

As explained in previous sections, the Scalar and Tensor sectors in the SVT generalizations of Horndeski theory will be identical to Horndeski theory. Let us only write the action for the graviton in the usual form \cite{Kobayashi:2019hrl}, which will be of use in the next section,
\[\label{tensor_action} S_{tensor}^{(2)}=\int \mathrm{d} t \mathrm{~d}^3 x\, a^3  \left[\mathcal{G}_\tau \left(\dot{h}_{i j}\right) ^2-\frac{\mathcal{F}_\tau}{a^2}\left( \overrightarrow{\nabla} h_{i j}\right)^2\right]\,,\]
where $\mathcal{G}_\tau$ and $\mathcal{F}_\tau$ give the speed of gravitational waves in the cosmological medium ($c_g$) as,
\begin{align}
c_g^2=\frac{\mathcal{F}_\tau}{\mathcal{G}_\tau}=\frac{2 G_4 - X \left(2 \ddot{\pi} G_{5X} + G_{5\pi}\right)}{2 G_4 + X \left(-4 G_{4X} - 2 H \dot{\pi} G_{5X} + G_{5\pi}\right)}\,.\label{eqn cg}
\end{align}
Let us recall that, as discussed before, the tensor sector of the $\mathcal{L}_{SVT}$ part of the action can be non-trivial on less symmetric backgrounds. 

Now, by a similar reasoning, since all mixed vector perturbations at quadratic order of the form $\delta g_{\cdot\, \mu}\, \delta A^\mu$ will be multiplied by a component of the background field $A^{(0)}=0$, then, the vector perturbations of the SVT theory $\mathcal{L}_{SVT}$ will not mix with the vector perturbations of Horndeski theory. Indeed, the complete vector sector in the unitary gauge takes the form 
\begin{flalign}
  S^{(2)}_{vector} = -\int &\mathrm{d}t\,\mathrm{d}^3x\,a^3\,   \left[\dfrac{1}{a^2} \mathcal{G}_{V}\: \left(\dot{{A}_i}\right)^2- \frac{1}{a^2} \mathcal{F}_{V}\, \left( \dfrac{1}{a} \partial_i{A}_j\right)^2 + \mathcal{K} \, \left( \dfrac{1}{a} \partial_i{e}_j\right)^2 \right]\,,\label{eqn vector sectorHSVT0}
\end{flalign}
for the gauge invariant combination of metric perturbations $e_{i} = \dot{W}_{i} - 2 H W_{i} + Z_{i}$. Thus, as always in Horndeski theory, the vector perturbations of the metric are non-dynamical, leaving only the non--trivial vector sector of the SVT generalization,
\begin{flalign}
  S^{(2)}_{vector} = -\int &\mathrm{d}t\,\mathrm{d}^3x\,a\,   \left[ \mathcal{G}_{V}\: \left(\dot{{A}_i}\right)^2- \frac{1}{a^2} \mathcal{F}_{V}\, \left( \partial_i{A}_j\right)^2  \right]\,,\label{eqn vector sectorHSVT}
\end{flalign}
where the speed square of the vector modes is 
\begin{eqnarray}
    c_A^2=\frac{\mathcal{F}_V}{\mathcal{G}_V}\,,
\end{eqnarray}
with
\begin{align}
    \mathcal{G}_{V} &=2\,\lambda_1+X \left(\lambda_5-\,\lambda_{6\pi}-2\,\lambda_{2\pi}\right)\nonumber\\
    &+H \dot{\pi}  \left(\left(2\,\lambda_{6}+8\,\lambda_{2}\right)+X  \left(4\,\lambda_{6X}+8\,\lambda_{2X}+4\,\lambda_{7}\right)\right)\nonumber\\
    &+{H}^{2\,} \left(-4\,\lambda_{10}-X \left(3\,\lambda_{3}+8\,\lambda_{10X}\right)-2{X}^{2\,} \lambda_{3X}\right)
\end{align}

\begin{align}
\mathcal{F}_V &= 2\,\lambda_1+X \left(2\,\lambda_{2\pi}+\lambda_{6\pi}\right)\\
&+H\, \dot{\pi}  \left(4\,\lambda_{2}+\lambda_{6}-X \,\lambda_{3\pi}\right)\nonumber\\
&-{H}^{2\,} \left(4\,\lambda_{10}+X    \lambda_{3}\right)\nonumber\\
&-\dot{H} \left(4\,\lambda_{10}+X    \lambda_{3}\right)\nonumber\\
&+\ddot{\pi}\left(4\,\lambda_{2}+\lambda_{6}+X \left(2\,\lambda_{7}+4\,\lambda_{2X}+2\,\lambda_{6X}\right)\right)\nonumber\\
&-2\ddot{\pi}\,H\, \dot{\pi}\left( \,\lambda_{3X}X+ \,\lambda_{3}+4\,\lambda_{10X}\right)\,.\nonumber
\end{align}

Let us note that the SVT lagrangians $L_{k\in \{8,9,11,12\}}$ are of the order $F^4$, and thus, they do not contribute at linear order on the cosmological background. Furthermore, let us note that by the same arguments as above, in the absence of proper Scalar--Tensor modes ---if we had not added the Horndeski part to the Lagrangian (\ref{eqn lagSVTFinal})--- then we would have obtained $\dot\pi=0$, thus also recovering the Maxwell action from (\ref{eqn vector sectorHSVT}). This suggests that in regions where the time variation of the scalar field $\pi$ is negligible ---and assuming a cosmological background--- the SVT generalization of Horndeski theory may reduce to Maxwell electrodynamics plus General Relativity (See also a discussion in the next section and in \cite{Babichev:2024kfo,Mironov:2024idn}). Furthermore, in general, due to the fact that the Scalar and Tensor sectors remain untouched by these SVT generalizations, all results for the stability of nonsingular solutions in Horndeski theory \cite{Libanov:2016kfc, Kobayashi:2016xpl, Mironov:2019fop,Volkova:2024mbn} and the Vainshtein screening follow as usual \cite{Vainshtein:1972sx,Babichev:2013usa,Kobayashi:2019hrl,Koyama:2013paa}.

\section{Family of Luminal SVT generalizations of Horndeski theory}\label{sec luminalSVT ext}

The separation of Scalar--Vector--Tensor perturbations between the pure SVT couplings in the part $\mathcal{L}_{SVT}$ of the total Lagrangian (\ref{eqn fullL}) and the Horndeski Lagrangian, at first order in the equations of motion, puts us in the very comfortable position to find whole families\footnote{Of course, it is expected that the theories can be distinguished one from another at higher order in perturbation theory or at linear order, but on less symmetric backgrounds.} of theories that satisfy a given physically compelling property, at leading order for the vector modes, while leaving unmodified the physically interesting solutions of Horndeski theory. 

For instance, the near equality of speed of gravitational waves and light $\vert \frac{c_g}{c_A}-1 \vert \leq 5 \times 10^{-16}$ \cite{LIGOScientific:2017vwq,Abdalla:2022yfr} was traditionally interpreted as constraining the modifications of gravity that change the speed of gravity away from unity, on any cosmological medium. Indeed, with a photon not coupled to the scalar mode ($c_A=1$), the experimental observations would probably require $c_g=1$, which in Horndeski theory would amount to take  $G_{4X}=G_{5\pi}=G_{5X}=0$ \cite{Bettoni:2016mij,Ezquiaga:2017ekz,Sakstein:2017xjx,Baker:2017hug,Creminelli:2017sry}. As can be seen from the expression (\ref{eqn cg}) for a solution independent of the matter content; namely, without using the background equations to express $H(G_2,G_3,\dots)$ and $\ddot{\pi}(G_2,G_3,\dots)$ in functions of $X$, to take them in the definitions $G_4$ and $G_5$. This would strictly  reduce the theory space of nonminimal couplings in Horndeski theory to Brans--Dicke type couplings to curvature, namely, only  $G_4(\pi)$. 

However, this direct interpretation of the bound $\vert \frac{c_g}{c_A}-1 \vert \leq 5 \times 10^{-16}$ was challenged in \cite{Mironov:2024idn,Mironov:2024wbx}. The different perspective proposed in these papers is to add couplings of the Photon to the Scalar mode, such that
\begin{equation}
    \frac{c_g(t)}{c_A(t)}=1\,,\label{eqn ratio1}
\end{equation}
holds in Horndeski theory, at all times, even with non--zero $G_{4X},\,G_{5\pi}$. In other words, the scalar--Photon couplings shown in \cite{Mironov:2024idn} are such that the speed of the Photon exactly tracks the speed of the graviton, such that they always satisfy  (\ref{eqn ratio1}).

More precisely, the Scalar--Photon couplings proposed in \cite{Mironov:2024idn}  take the form,
\begin{align}
    \mathcal{L}_{4A}=&- \tfrac{1}{4} G_4 F_{\alpha \beta } F^{\alpha \beta } + G_{4X}\, F_{\alpha }{}^{\delta } F_{\beta \delta } \nabla^{\alpha }\pi \nabla^{\beta }\pi \,,\label{eqn G4A}\\
    \nonumber\\
    \mathcal{L}_{5A}=&\tfrac{1}{8} G_5 \bigl(- F_{\alpha \beta } F^{\alpha \beta } (\nabla_{\gamma}\nabla^{\gamma}\pi) + 4 F_{\alpha }{}^{\delta } F_{\beta \delta } \nabla^{\beta }\nabla^{\alpha }\pi - 4 F_{\alpha }{}^{\beta } \nabla^{\alpha }\pi \nabla_{\delta }F_{\beta }{}^{\delta }\bigr)\,. \label{eqn G5A}
\end{align}
They appear in the action with fixed Lagrangian functions that match those of Horndeski theory $G_4$ and $G_5$. Thus they are very specific SVT couplings that when added to the Lagrangian of Horndeski, satisfy (\ref{eqn ratio1}).

It is interesting to note that by constraining the Lagrangian for the modified Photon on cosmological scales (\ref{eqn G4A}) and (\ref{eqn G5A}), one would immediately constrain the Lagrangian of Horndeski, $G_4,\,G_5$. Which may be relevant in the new era of multi--messenger astronomy. Indeed, modifications of gravity could be in principle  distinguished between one and another in an indirect manner, by looking at other tests that probe (\ref{eqn G4A}) and (\ref{eqn G5A}) directly, instead of gravitational tests for (\ref{eqn Horndeski}).

However, this strong link between the SVT and Horndeski actions can also be a disadvantage, because, although these theories could be used for late time cosmology---identifying $\pi$ with the scalar of Dark Energy--- the precise linearity of the Photon is also essential for many other cosmological phenomena. Besides, there are already  experimental constraints on disformally coupled scalars to the Photon, on which the Lagrangian (\ref{eqn G4A}) classifies (See for instance \cite{Brax:2014vva, Brax:2015hma,Babichev:2024kfo}). To the latter, we show in this work that the Lagrangians (\ref{eqn G4A}) and (\ref{eqn G5A}) are just special cases of a family of SVT generalizations of Horndeski theory, which also align with the experimental constraint on the ratio of speeds, by exactly satisfying (\ref{eqn ratio1}). This, while introducing extra freedom in the SVT couplings, which may be useful for independent modifications of the tensor and vector sectors, while still passing the speed test.

More precisely, the class of SVT theories with second order equations of motion, with the action (\ref{eqn fullL}) and the following  constraints in the scalar functions can satisfy (\ref{eqn ratio1})
\begin{subequations}
    \begin{align} 
&\lambda_{10} =  - \frac{1}{4}\,X \lambda_{3},\label{eqn luminalConstraints1} \\
&\lambda_{7} = 2\,\lambda_{2X} - \frac{3}{2}\,\lambda_{3\pi}-\,X \lambda_{3X \pi}, \\
&\lambda_{6} = -4\,\lambda_{2}+X \lambda_{3\pi}, \\
&\lambda_5 = \frac{4\,\left(\lambda_1 \left(-2\,G_{4X}+G_{5\pi}\right)+\left({G_{4}}-\,X G_{4X}\right) \left(-2\,\lambda_{2\pi}+X \lambda_{3\pi\pi}\right)\right)}{\left(2\,{G_{4}}-\,X G_{5\pi}\right)}\,,\label{eqn luminalConstraints4} \end{align}
\end{subequations}

Indeed with the constraints (\ref{eqn luminalConstraints1})-(\ref{eqn luminalConstraints4}), the quadratic action for the vector modes of the SVT Lagrangians (\ref{eqn vector sectorHSVT}) takes the final form
\begin{flalign}
S^{(2)}_{vector}= - \int &\mathrm{d}t\,\mathrm{d}^3x\,a\:  f(\lambda)\, \left[ c_g^{-2}\: \left(\dot{{A}_i}\right)^2- \frac{1}{a^2} \, \left( \partial_i{A}_j\right)^2  \right]\,,\label{eqn vector sectorHSVT Luminal}
\end{flalign}
with 
\begin{equation}
    f (\lambda)=2\, \lambda_1-2\lambda_{2\pi}\,X+\lambda_{3\pi\pi}\,X^2\,.
\end{equation}
Therefore, even with the time varying function $f(\lambda)=f(\pi,\,X)$, the speed of the vector modes $c_A^2=1/c_g^{-2}$ exactly tracks the speed of Gravitational waves (\ref{eqn ratio1}). These theories contain the actions (\ref{eqn G4A}) and (\ref{eqn G5A}) obtained by a Kaluza--Klein reduction in \cite{Mironov:2024idn} as a special case, yet there is additional freedom through $f(\pi,\,X)$ in the vector sector.

Let us note that 
\begin{equation}
    f(\lambda)\:\overset{X \rightarrow 0}{\longrightarrow}\: 2\,\lambda_1\,,
\end{equation}
so that as commented before, when the time variation of the scalar of Dark Energy $\dot\pi$ is negligible, and when $\lambda_1$ is constant, we recover Maxwell electrodynamics.

\section{Conclusions}\label{sec conclusions}

We analyzed generalizations of Kaluza--Klein compactifications of 5D Horndeski theory. They take the form of general Scalar--Vector--Tensor theories,  where the higher derivative vector field is U(1)-gauge invariant and the Scalar--Tensor sector corresponds to Horndeski theory. The equations of motion remain of second order. The theory is written in terms of additional free functions of the scalar field $\pi$ and $X$  besides the 4 functions of Horndeski theory. They contain Lagrangians of up to second power in second derivatives of the scalar and up to four powers of the U(1) vector field strength, such that they contain the Kaluza--Klein reduction of 5D Horndeski theory as a special case.

Then, we analyzed the subset of SVT generalizations of Horndeski theories such that the speed of the tensor and vector modes is exactly  the same on any cosmological background, even for the general $G_4(\pi,X)$ and $G_5(\pi)$ functions in the Horndeski action. The complete theory retains the freedom of the four Horndeski functions in the Scalar--Tensor sectors, plus new freedom in the vector sector, through an independent  function $f(\pi,\,X)$ on the FLRW background. The latter may be useful for independent modifications of the tensor and vector sectors, while the speed of the vector modes still tracks the speed of the tensor modes. 

We also found that with these SVT theories it is not possible that the speed of the vector modes adjusts to that of the Horndeski graviton with nonzero $G_{5X}$. Thus, $G_5(\pi,X)$ is still restricted by the $c_g=c_A$ constraint.

Also, it would be interesting to extend this class of SVT theories free of ghosts by exploring degeneracy conditions. We leave that for further studies.

\section*{Acknowledgements}

The authors are thankful to P. Petrov for useful discussions.

The work on this project has been supported by Russian Science Foundation grant № 24-72-10110,

\href{https://rscf.ru/project/24-72-10110/}{  https://rscf.ru/project/24-72-10110/}.

\bibliographystyle{IEEEtran}
\bibliography{main}

\end{document}